\documentstyle[a4wide,epsf,11pt,titlepage]{article}

\newcounter{nref}
\newcommand{\bbib}{%
  \renewcommand{\refname}{\large\bf References}%
  \begin{thebibliography}{99}%
  \setcounter{enumiv}{\thenref}}
\newcommand{\ebib}{%
  \end{thebibliography}%
  \setcounter{nref}{\arabic{enumiv}}}
\newcommand{\head}[3]{%
  \setcounter{nref}{0}%
  \thispagestyle{empty}%
  \section*{\LARGE\bf #1}%
  \stepcounter{section}%
  \addcontentsline{toc}{section}{#1}%
  \large\itshape%
  #2\\\vspace{0.1pt}\\%
  #3%
  \normalsize\upshape%
  \bigskip}

\begin{document}


\head{Energy exchange inside SN ejecta \\ and light curves of SNe~Ia}
     {E.I.\ Sorokina$^1$, S.I.\ Blinnikov$^{2,1}$}
     {$^1$ Sternberg Astronomical Institute, Moscow, Russia\\
      $^2$ Institute for Theoretical and Experimental Physics, Moscow, Russia}

A treatment of line opacity is one of the most crucial problems
for the light curve (LC) modeling of Type Ia supernovae (SNe~Ia).
Spectral lines are the main source of opacity inside SN~Ia ejecta
from ultraviolet through infrared range.
A lot of work has been done on this subject
\cite{sorokina.Ba,sorokina.Bl,sorokina.EK,sorokina.EP,sorokina.KLCS,%
sorokina.WPV}.

The problem itself can be divided into two parts:
\begin{itemize}
\item How flux and energy equations are changed in the expanding medium;
\item How one should average flux and intensity, as well as extinction
and absorption coefficients, over energy bins used in calculations.
\end{itemize}

Almost all previous research was about the flux equation.
Here we will focus on the energy equation.
We will suppose a free expansion of gas, i.e. linear law
for the velocity distribution $v=r/t$.
The principal difference between the behavior of flux and intensity
in comparison to the static case is that the flux always
becomes lower in the expanding media, while the sign of the change
of intensity depends on the temperature gradient.
It is clear qualitatively.
When we are sitting inside the gas and would like to measure a flux
and an intensity at a fixed frequency in the static case
we should just measure how the surrounding gas emits and absorbs
at this frequency.
If this specific frequency corresponds to a strong line we will  not be able
to see distant layers of gas, since rather large optical depth accumulates
quite close to the observer.
In the continuum we can see rather distant gas layers.
So at the frequency of a strong line we observe almost local gas
without temperature gradient, hence we measure rather low flux value,
and the intensity corresponds to the local blackbody value.
In the continuum the flux grows appreciably because of growing gradients,
and the intensity corresponds to the blackbody at $\tau \sim 1$,
where conditions differ from local ones.

\begin{figure}[ht]
  \centerline{\epsfxsize=0.45\textwidth\epsffile{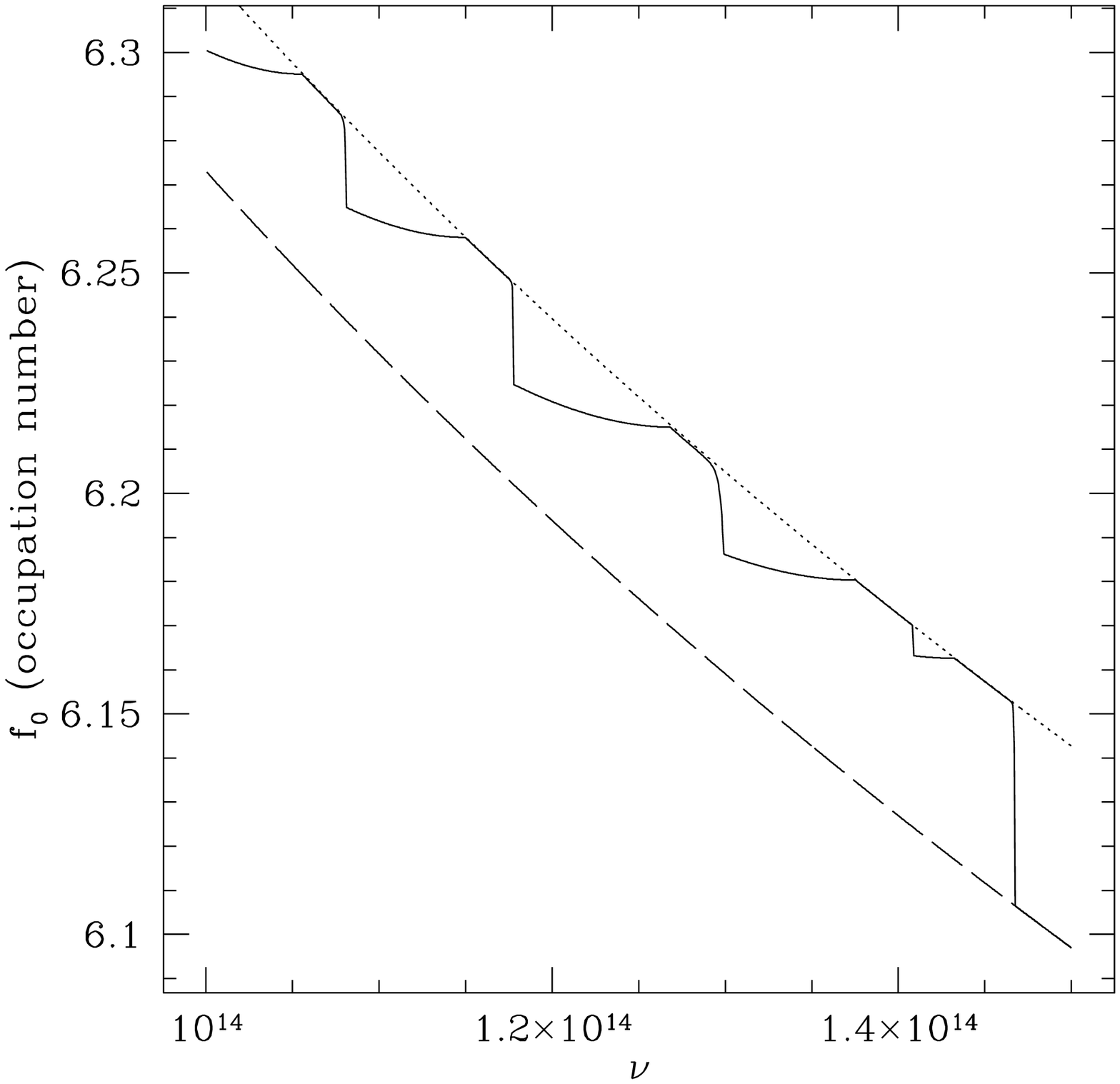}
              \epsfxsize=0.45\textwidth\epsffile{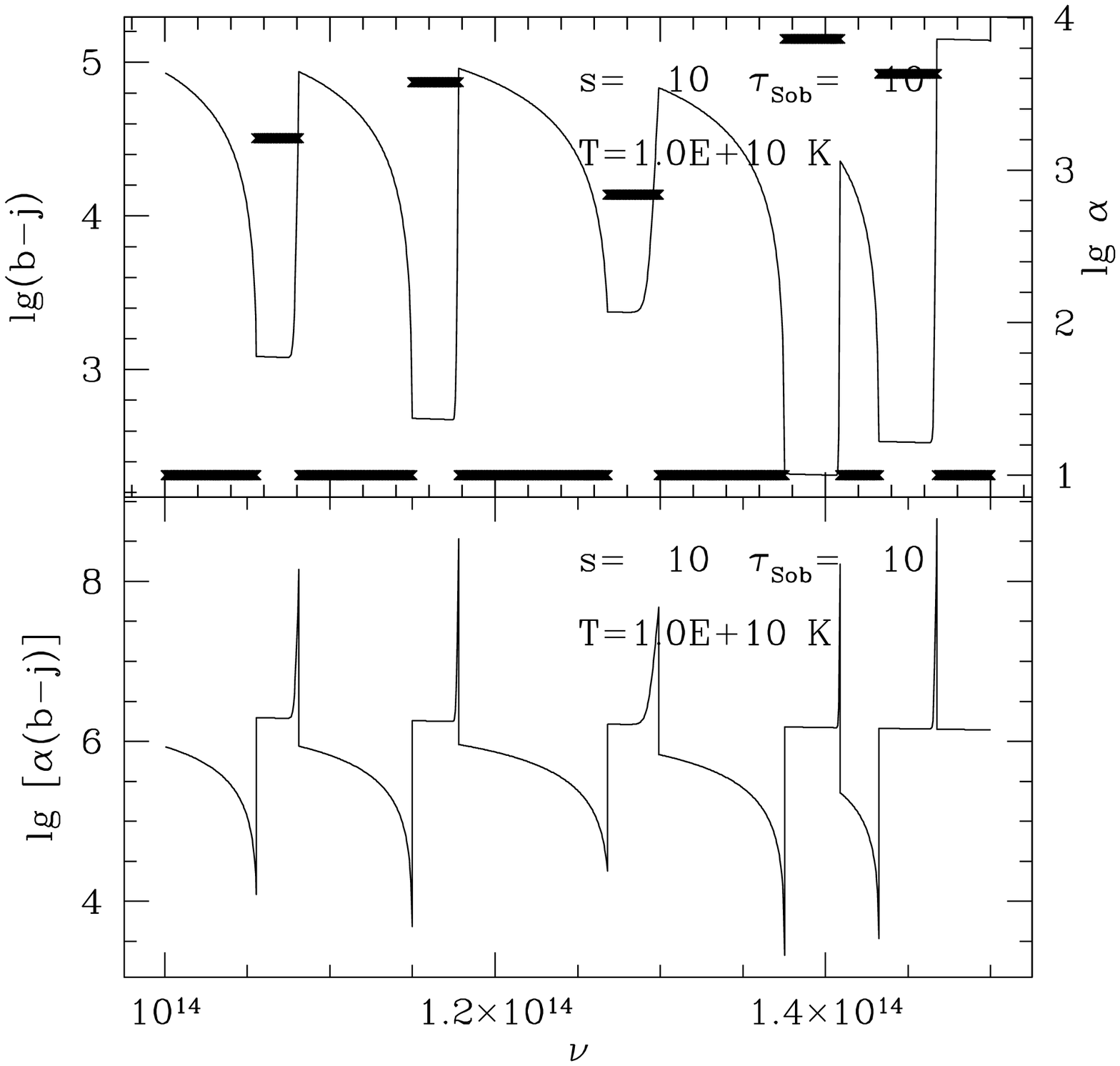}}
  \caption{{\it Left panel: } the zero moment of an occupation number $f_0$.
Dots correspond to the local blackbody value, dashes show the blackbody
at the distance where continuum forms, solid line demonstrates the behavior
of $f_0$ when five spectral lines are present in the frequency interval we are
interested in.
{\it Right panel: } logarithms of monochromatic $\alpha$ (thick lines,
``top hat'' profile), $b-j$, and $\alpha(b-j)$.
The main problem is to integrate the last function, especially
when the number of lines is not 5, but several hundreds.}
  \label{sorokina.figf0}
\end{figure}

In the expanding case, a photon's mean free path becomes less than in the
static medium,
even at the frequency which corresponds to the continuum
at the rest observer frame.
If we encounter a strong line blueward, before $\tau \sim 1$ is reached
in continuum,
then, most probably, we will not be able to see further out.
So the temperature gradient at the mean free path of a photon becomes less
than in the static case, therefore the flux drops down.
The intensity in the expanding medium would then correspond to the blackbody
radiation at the distance where the line forms, but it is redshifted
by Doppler effect.
For a constant temperature and pure continuum, the intensity
at any space point and any frequency is lower
than the local blackbody intensity.
A strong nearby line enhances the value of the observed intensity
(opposite to the behavior of the flux, which drops down with the comparison
to a pure continuum case) and diminishes the difference
between this intensity and a local blackbody.
Fig.~\ref{sorokina.figf0} shows all these dependencies for the zero moment
of an occupation number, which is proportional to an angle averaged intensity
(the zero moment of intensity): $J_\nu={2h\nu^3 \over c^2}f_0$;
$J_\nu={1 \over 2}\int_{-1}^1 I_\nu d\mu$.

To derive these dependencies formally, we have to solve the Boltzmann equation
in the comoving frame for a spherically-symmetrical flow:
\begin{equation}
 {1\over c}{\partial f\over \partial t} + \mu{\partial f\over \partial r}
+ {1-\mu^2 \over r}{\partial f\over \partial \mu}
-{\nu \over ct}{\partial f\over \partial \nu} =
  \eta_\nu - \chi_\nu f \; ,
\end{equation}
where we put $r=vt$, which is quickly established in
SN ejecta.
When we apply a diffusion limit, so that the space derivatives are negligible,
the solution of the Boltzmann equation is~\cite{sorokina.Bl}
\begin{equation}
  f_0 =
  \int_\nu^{t\nu/t_0}
  { \nu \over \tilde\nu^2 }
 \;
 \alpha^\star (t\nu/\tilde\nu, \tilde\nu) \,
 b(t\nu/\tilde\nu, \tilde\nu)
  \exp\left[ - \int_\nu^{\tilde\nu}
  { \nu \over \bar\nu^2 }
   \; \alpha^\star (t\nu/\bar\nu, \bar\nu) \; d\bar\nu  \right] \;
 d\tilde\nu \; ,
 \label{sorokina.f0}
\end{equation}
where $b$ is a blackbody occupation number, and the dimensionless
absorption coefficient $\alpha^\star = \alpha ct$.

\begin{figure}[ht]
  \centerline{\epsfxsize=0.6\textwidth\epsffile{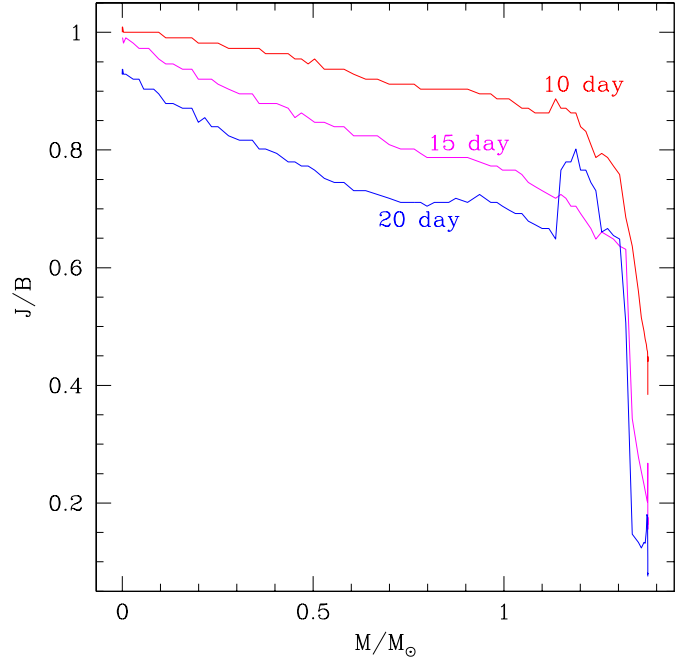}}
  \caption{The distribution of $J/B$ inside the ejecta at different times
demonstrates how matter and radiation are decoupled.}
 \label{sorokina.figJBM}
\end{figure}

Last equation gives us an exact solution for a monochromatic value of $f_0$.
Numerical codes which are used for computations of SNe radiative evolution
operate with averaged numbers, and each energy bin contains hundreds
of spectral lines.
During such computations one needs to solve equations for energy exchange
between radiation and gas in each bin:
\begin{equation}
 {d <j_\nu> \over dt } \; \propto \; <\alpha_\nu>(<b_\nu>-<j_\nu>).
\end{equation}
The problem here is to average $\alpha$ correctly inside a bin with hundreds
(or thousands) of lines.
The right way is to average it so that
\begin{equation}\label{sorokina.avgalp}
 <\alpha>={<\alpha(b-f_0)> \over <b>-<f_0>} \qquad (f_0 \equiv j_\nu).
\end{equation}

The main difficulty here is to derive an average value of $<\alpha(b-f_0)>$.
As we said  above, in the strong lines $f_0=b$.
If an optical depth in the continuum is also very high, so that radiation is
tightly coupled with matter and $f_0=b$, then $b-f_0 \approx 0$ and it is not
so important how we treat $\alpha$.
It becomes important when radiation is decoupled from matter.
As it is seen in the fig.~\ref{sorokina.figJBM}, they are decoupled
within the entire SN~Ia ejecta even before maximum light (which occurs about 20th day
after explosion).
A similar result was obtained in~\cite{sorokina.EPThn}.
So the correct treatment of $<\alpha>$ is essential.
We need to take the integral~(\ref{sorokina.f0}) accurately.
It is very hard to do numerically, since $\alpha$ jumps up and down
by many orders of magnitude hundreds of times.
Moreover, where $\alpha$ is high, there $b-f_0$ is extremely  small
(see fig.~\ref{sorokina.figf0}).
Analytically, one can solve this problem, but only for very simple line profiles.
We assume that lines have a rectangle shape (``top hat'' profile).
For rectangle lines in the Rayleigh--Jeans regime and infinite expanding
medium (so that at given $\nu$ one can observe emission with arbitrary
shifted $\Delta\nu$ in the transparent case) we have

\begin{eqnarray}
   f_0(\nu) = { kT \over h\nu}
   \Biggl[\!\!\Biggl[
    1 - {1\over\alpha^\star_{N_\nu}}                 
           \left( 1 - e^{-\alpha^\star_{N_\nu}(1-{\nu\over\nu_{N_\nu}})}
           \right)  
  \qquad   \qquad \qquad \qquad \qquad \qquad \qquad \qquad    \nonumber\\
%
      - \sum_{j=N_\nu+1}^{N_{\rm max}} \left[
          {1\over\alpha^\star_j}
             \left( 1 - e^{-\alpha^\star_j ( {\nu\over\nu_{j-1}}
                                                - {\nu\over\nu_{j}} )}
             \right)
          \exp\left\{
         -\sum_{i=N_\nu}^{j-1} \alpha^\star_i\left({\nu \over \nu_{i-1}}
          - {\nu \over \nu_i}\right) \right\}
                                       \right]
        \qquad   \nonumber  \\
%
    {}  - {1\over\alpha^\star_{N_{\rm max}+1}}
           \left( 1 - e^{-\alpha^\star_{N_{\rm max}+1}\nu/\nu_{N_{\rm max}}}
           \right) \exp\left\{
         -\sum_{i=N_\nu}^{N_{\rm max}} \alpha^\star_i\left({\nu \over \nu_{i-1}}
          - {\nu \over \nu_i}\right) \right\}
   \Biggr]\!\!\Biggr]
\end{eqnarray}

To get $<\alpha>$ from~(\ref{sorokina.avgalp}), we need to integrate
this expression once more.
This also can be done analytically, and adds the third summation
over all lines in the bin.
Computing of these long sums is very time consuming.
Probably, the efficiency can be improved along the way proposed
in~\cite{sorokina.BWW}.

\begin{figure}[ht]
  \centerline{\epsfxsize=0.6\textwidth\epsffile{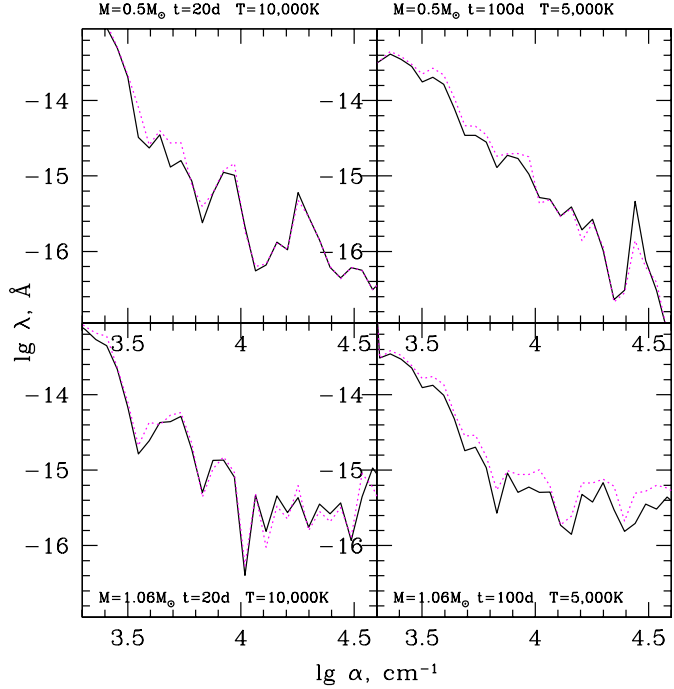}}
  \caption{Logarithm of absorption coefficient calculated accordingly
to our formulae (solid) and in Eastman--Pinto approximation (dotted).
Top panels show the absorption for two sets of (time, temperature)
inside the layer of W7 model which consists of mostly ${}^{56}$Ni,
while bottom panels correspond to the layer of intermediate mass elements.
}
  \label{sorokina.figalp4}
\end{figure}

Using our expression for $<\alpha>$, we have computed the opacity tables
for standard SN~Ia model W7.
The results are compared with the absorption derived
with the Eastman--Pinto approximation~\cite{sorokina.EP}
(see fig.~\ref{sorokina.figalp4}).
In the UV region, where free--free and bound--free emission dominates,
$<\alpha>$ remains almost unchanged, while it differs up to a factor of two
in the range between near UV through IR.

\begin{figure}[ht]
  \centerline{\epsfxsize=0.6\textwidth\epsffile{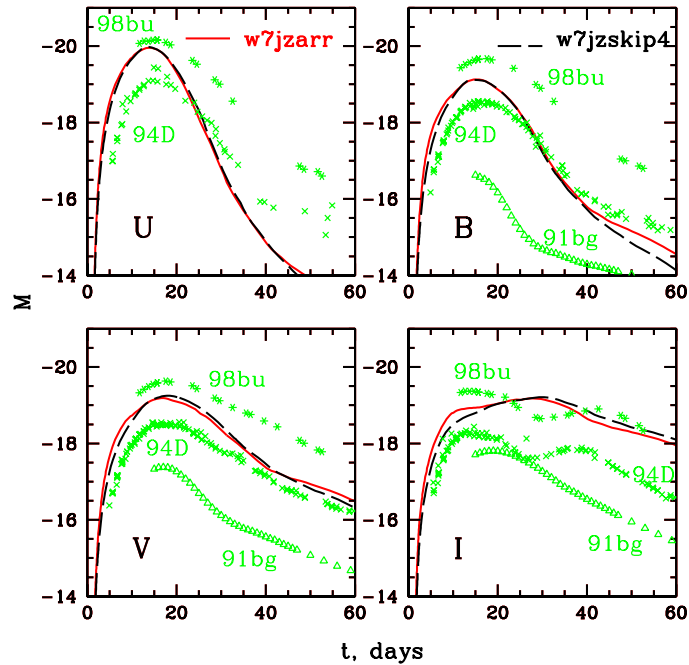}}
  \caption{UBVI light curves of the model W7 with two opacity approximations:
Eastman--Pinto (black dashed) and the way of averaging presented here (red solid).
UBV bands remain almost unaffected, while I--band shows now
a complicated structure, more similar to what one can observe for real SNe~Ia,
though not identical.
Crosses, stars and triangles show the observational LCs for three SNe~Ia.}
  \label{sorokina.figLC}
\end{figure}

LCs for W7 (fig.~\ref{sorokina.figLC}) were calculated with the help
of the code STELLA~\cite{sorokina.BEBPW}.
UBV LCs remain almost unchanged, while I-band
is affected appreciably (fig.~\ref{sorokina.figLC}).
This is explained by the changes of opacity and spectrum in the I-band.

\subsection*{Acknowledgements}

We are grateful to the organizers of the Ringberg meeting
for their hospitality, to Stan Woosley for supporting our work in the UCSC.
Our work in US was supported by grants NSF AST-97 31569 and
NASA - NAG5-8128, in Russia RBRF 99-02-16205 and RBRF 02-02-16500.


\end{document}